\documentclass[10pt,prl,twocolumn,floatfix,tightenlines,showpacs,showkeys,preprintnumbers,altaffilletter, nofootinbib, superscriptaddress, longbibliography]{revtex4-1}
\usepackage{graphicx,mathtools,tabu}
\usepackage{epstopdf}
\usepackage{amsmath}
\usepackage{amsfonts}
\usepackage[colorlinks=true,citecolor=red,linkcolor=blue,breaklinks=true]{hyperref}
\usepackage[dvipsnames]{xcolor}
\usepackage{accents}
\usepackage[figure, table]{hypcap}
\usepackage{amssymb}
\usepackage{appendix}
\usepackage{comment}
\usepackage{bbold}
\usepackage{color}
\usepackage{slashed}
\usepackage{subfigure}
\usepackage{setspace}
\usepackage{footnote}
\usepackage{multirow}
\usepackage{braket}
\usepackage[normalem]{ulem}
\usepackage[utf8]{inputenc}
\usepackage{mathrsfs}
\usepackage{longtable}
\usepackage{enumitem}
\usepackage{orcidlink}  

\def\dm{\delta m^2}
\def\Gpc{{\rm Gpc}}
\def\Mpc{{\rm Mpc}}

\def\eV{{\rm eV}}

\def\TeV{{\rm TeV}}
\def\m{{\rm m}}

\date{\today}

\begin{document}
\preprint{}
\title{Constraints on pseudo-Dirac neutrinos using high-energy neutrinos from NGC 1068 }

\author{Thomas Rink\,\orcidlink{0000-0002-9293-1106}}
\email{thomas.rink@mpi-hd.mpg.de}
\affiliation{Max-Planck-Institut f{\"u}r Kernphysik, Saupfercheckweg 1, 69117 Heidelberg, Germany}

\author{Manibrata Sen\,\orcidlink{0000-0001-7948-4332}}
\email{manibrata@mpi-hd.mpg.de}
\affiliation{Max-Planck-Institut f{\"u}r Kernphysik, Saupfercheckweg 1, 69117 Heidelberg, Germany}

\begin{abstract}
\noindent Neutrinos can be pseudo-Dirac in Nature - they can be Majorana fermions while behaving effectively as Dirac fermions. Such scenarios predict active-sterile neutrino oscillations driven by a tiny mass-squared difference $(\delta m^2)$, which is an outcome of soft lepton number violation. Oscillations due to tiny $\delta m^2$ can only take place over astrophysical baselines and hence are not accessible in terrestrial neutrino oscillation experiments. This implies that high-energy neutrinos coming from large distances can be naturally used to test this scenario. We use the recent observation of high-energy neutrinos from the active galactic nuclei NGC 1068 by the IceCube collaboration to rule out $\delta m^2$ in the region $[1.4 \times 10^{-18}, 10^{-17}]\, {\rm eV}^2$ at more than $90\%$ confidence level - one of the strongest limits to date on the values of $\delta m^2$. We also discuss possible uncertainties which can reduce the sensitivity of these results.
\end{abstract}
\maketitle

\noindent\textbf{\emph{Introduction -- }}The advent of multi-messenger astronomy in the last decade has been one of the biggest success stories of particle physics~\cite{2017muas.book.....B}. The observation of high-energy neutrino events, in direct correlation with the gamma-ray flares from the blazer TXS $0506+056$, identified blazars as powerful sources of these neutrinos~\cite{IceCube:2018cha,IceCube:2018dnn}. Since then, additional searches have been performed to locate more of these sources. The observations of these high-energy neutrinos, which can point back to their sources, indicate the presence of some of the most powerful cosmic accelerators in the Universe.

Very recently, the IceCube collaboration released a search result for neutrino emission from a list of 110 gamma-ray sources during the period 2011-2020~\cite{IceCube:2022der}. An excess of $79^{+22}_{-20}$ neutrino events, with energies of few TeV, was reported to be coming from the direction of a nearby active galactic nucleus (AGN) -- the NGC 1068 -- with a global significance of $4.2\sigma$, which was a major improvement over a previous result of $2.9\sigma$. The neutrino spectra can be fit using a power-law, $\Phi_{\nu_{\mu} + \bar{\nu}_{\mu}} = \Phi_{0} \left( E_{\nu}/\mathrm{1\,TeV}\right)^{-\gamma}$, with $\Phi_{0}$ being the overall flux normalization at neutrino energy $E_{\nu}=\mathrm{1\,TeV}$ and $\gamma$ the spectral index. 
Using this, the collaboration reported $\Phi_{0}^{\mathrm{1TeV}} = (5.0 \pm 1.5_{\mathrm{stat.}} \pm 0.6_{\mathrm{sys.}})\times 10^{-11}\,  \mathrm{TeV}^{-1} \mathrm{cm}^{-2} \mathrm{s}^{-1}$, with $\widehat{\gamma}=3.2 \pm 0.2_{\mathrm{stat.}}\pm 0.07_{\mathrm{sys.}}$. The distance to the AGN is estimated to be $d=14.4\,\Mpc$, although it has been quoted previously in the literature to lie between $10.3\pm 3\,\Mpc$~\cite{1992pngn.conf.....T} and $d=16.5\,\Mpc$~\cite{Traianou:2019sxz}.

The discovery of these high-energy neutrinos has opened up new avenues of testing exotic physics, which was otherwise inaccessible in terrestrial laboratories. The arrival of these neutrinos clearly indicates that the Universe is not opaque to neutrinos at this energy range. This can be used to put tight constraints on different kinds of non-standard physics in the new sector (see~\cite{Kelly:2018tyg} and references therein for a detailed discussion). 

The naturally long baseline offered by these neutrinos allows us to test the violation of lepton number in the Standard Model (SM). If lepton number is not considered a symmetry of the SM, and we know that it is already broken by quantum effects even within the SM, neutrinos can be Majorana fermions. The extent of lepton number violation in the SM is quantified by the Majorana mass term of neutrinos in the Lagrangian. However, if lepton number is violated \emph{softly} - measurable through a tiny Majorana mass term - neutrinos can be pseudo-Dirac, where they behave effectively as Dirac neutrinos~\cite{Wolfenstein:1981kw,Petcov:1982ya,Bilenky:1983wt,Kobayashi:2000md,Anamiati:2017rxw,deGouvea:2009fp,Vissani:2015pss}\footnote{Note that historically, the term ``pseudo-Dirac'' was used to denote a pair of active neutrinos behaving as a Dirac particle, whereas the term ``quasi-Dirac'' was used for an active-sterile pair. However, nowadays the term ``pseudo-Dirac neutrinos'' refer to an active-sterile neutrino pair.}. In this case, the neutrino mass-eigenstates corresponding to the active and sterile states develop a \emph{tiny} mass-squared difference, proportional to the extent of lepton number violation. Oscillations driven by this tiny mass-squared difference $(\dm)$ will only be relevant for long baselines, inversely proportional to $\dm$.

Clearly, for testing these oscillations due to tiny $\dm$, one needs access to long baselines. Hence, constraints on the pseudo-Dirac hypothesis from terrestrial experiments are weak~\cite{Das:2014jxa,Hernandez:2018cgc,Anamiati:2019maf}. Stronger constraints are available from atmospheric neutrinos~$\dm\lesssim 10^{-4}\eV^2$~\cite{Beacom:2003eu}, solar neutrinos~$\dm\lesssim 10^{-11}\eV^2$~\cite{deGouvea:2009fp,Ansarifard:2022kvy}, as well as supernova neutrinos~$\dm\lesssim 10^{-20}\eV^2$~\cite{deGouvea:2020eqq,Martinez-Soler:2021unz}, which naturally provide a longer baseline for neutrinos. This naturally opens up the possibility of using high-energy neutrinos coming from the far reaches of the Universe to probe even smaller values of $\dm$. Using models of gamma-ray bursts as sources for the astrophysical neutrino flux observed by IceCube, limits have been placed on values of $10^{-18}\eV^2\lesssim \delta m^2\lesssim 10^{-12}\eV^2$~\cite{Beacom:2003eu,Keranen:2003xd,Esmaili:2009fk,Esmaili:2012ac,Joshipura:2013yba,Brdar:2018tce}. Future observations of the cosmic neutrino background can also yield promising results for constraining tiny values of $\dm$~\cite{Perez-Gonzalez:2023llw}.

In this work, we utilize the high-energy neutrino events observed from NGC 1068 to probe neutrino oscillations due to a small $\dm$. The advantages are two-fold: firstly, there exists a precise measurement of the location of the source $d\sim 14.4\,\Mpc$, RA (right ascension) and DEC (declination) $ = (40.667, -0.0069) $; secondly, the neutrino energies are well measured to be around a few TeV to tens of TeV. The combination of these allows us to precisely pin down the oscillation length, and study the sensitivity of neutrino oscillations to tiny mass-squared differences. This allows us to rule out $\dm = 1.4\times 10^{-18}\eV^2$ at more than $90\%$ confidence level. We emphasize that this is the smallest value of $\dm$ constrained with such a high significance.  
An earlier work by one of the authors used data from SN1987A to probe even tinier values of $\dm\lesssim 10^{-20}\eV^2$~\cite{Martinez-Soler:2021unz}, however, the significance was lower due to the small number of events observed. This observation by IceCube, along with the data from SN1987A, provide two of the tightest constraints on the strength of lepton number violation in the SM.

\noindent\textbf{\emph{Pseudo-Dirac neutrinos -- }}
To explain the masses of the neutrinos, the SM can be expanded to include three additional sterile neutrinos with at least two of them having Majorana masses $M_N$, which characterize the extent of lepton number violation. Apart from the Majorana mass term, the neutrino mass matrix also consists of Dirac-type masses given by $M_D=Yv$, where $Y$ is the Yukawa coupling and $v$ denotes the vacuum expectation value of the Higgs field. In the limit of soft lepton number violation, $|M_N|\ll |Y v|$, the generic neutrino mass matrix can be diagonalized by a $6\times 6$ block-diagonal unitary matrix $U$, consisting of the PMNS matrix, and another $3\times 3$ unitary matrix~\cite{Kobayashi:2000md}. A generalization using the spectral density function was recently worked out in~\cite{Banks:2022gwq}.

For a tiny $|M_N|$, the mixing between the active and sterile states becomes maximal. In this case, a neutrino flavor state can be written as a superposition of two almost degenerate mass states,
\begin{align}
    \nu_{\alpha L}&=U_{\alpha k}\frac{(\nu_{k}^+ +i\,\nu_{k}^-)}{\sqrt{2}}\, ,
\end{align}
with masses $m_{k, \pm}^2 = m_k^2 \pm \delta m_k^2/2$. Here $\nu_{k}^\pm$ correspond to the mass eigenstates associated with the flavor state. This scenario leads to active-sterile oscillations driven by $\delta m_k^2/2$, in addition to oscillations due to solar $(\Delta m^2_{\rm sol})$ and atmospheric $(\Delta m^2_{\rm atm})$ mass-squared differences~\cite{Tanabashi:2018oca}. For simplicity, we will assume that $\dm$ is the same for all states, and hence drop the index $k$. 
The analysis with different $\delta m^2$ becomes considerably more complicated because of multiple parameters. In that case, we expect degeneracies to arise among different $\delta m^2$ when trying to fit the data within the (3\,+\,3) model. In particular, for IceCube-like experiments, which are mostly sensitive to muon neutrinos, we expect the sensitivity to be largest to $\delta m_{2}^2$ and $\delta m_{3}^2$, whereas for experiments sensitive to solar and supernova neutrinos (which mainly detect electron neutrinos and antineutrinos), the sensitivity should be mostly to  $\delta m_{1}^2$. Detailed investigation on this is need in the future.

For neutrinos travelling over astrophysical baselines, flavor oscillations induced by $(\Delta m^2_{\rm sol, atm})$ average out due to rapid oscillations, causing decoherence~\cite{Tanabashi:2018oca}. However, for small enough $\dm$, active-sterile oscillations can persist. The corresponding active neutrino survival probability can be computed as~\cite{deGouvea:2020eqq}
\begin{align}\label{eq:PDProb}
    P_{\alpha\alpha}(E_\nu) = \frac{1}{2}\left(1+e^{-\left(\frac{L}{L_{\rm coh}}\right)^2}\cos\left(\frac{2\pi L}{L_{\rm osc}}\right)\right)\,.
\end{align}
Here $L_{\rm osc}$ denotes the oscillation length and can be computed for a given neutrino energy $E$ and $\dm$ as,
\begin{equation} \label{eq:Losc}
    L_{\rm osc} = \frac{4\pi E_\nu}{\delta m^2}
    \approx 15\,\Mpc\left(\frac{E_\nu}{1\,\TeV}\right)\left(\frac{5\times 10^{-18}\,\eV^2}{\delta m^2}\right)\,.
\end{equation}
However, even these oscillations driven by $\dm$ can get washed out due to the separation of wave packets over long distances, leading to decoherence. This is measured through the coherence length $L_{\rm coh}$, which gives the length scale up to which flavor coherence is expected to be maintained. This depends also on the width of the neutrino wave-packet $\sigma_x$, which is usually determined from the process of neutrino production~\cite{Kersten:2015kio}. The coherence length is estimated as,
\begin{align}\label{eq:Lcoh}
    L_{\rm coh} &= \frac{4\sqrt{2} E_\nu^2}{|\delta m^2|}\sigma_x\, ,\\
     &\approx 10^6\,\Gpc \left(\frac{E_\nu}{1\,\TeV}\right)^2\left(\frac{5\times 10^{-18}\,\eV^2}{\delta m^2}\right)\left(\frac{\sigma_x}{10^{-10}\,{\rm\ m}}\right)\,.\notag
\end{align}
Taking into account the active sterile oscillation probability, and the flavor averaging due to the solar and atmospheric terms, the probability of obtaining a neutrino flavor $\nu_\beta$, starting from $\nu_\alpha$ is given by
\begin{align}\label{eq:Prob}
    P_{\alpha\beta} = P_{\alpha\alpha}(E_\nu; L, \dm)\sum_{k} \left|U_{\alpha k}\right|^2 \left|U_{\beta k}\right|^2 \,.
\end{align}
This is the new physics that we will constrain through our analysis in the next section. The long baseline $(d=14.4\,\Mpc)$ offered by the high-energy neutrinos observed by IceCube can be used to constrain tiny values of $\dm\sim 10^{-18}\eV^2$, which are otherwise difficult to access in other sources. 

\noindent\textbf{\emph{Analysis -- }}
The investigation performed in this work closely follows the procedure applied in Ref.~\cite{IceCube:2022der}.
Thus, we use an unbinned likelihood ratio method with a corresponding likelihood function defined as~\cite{Braun:2008bg}
\begin{align}\label{eq:LH}
        \log \mathcal{L} = \sum_{i} \log \left( \frac{n_{s}}{N} \mathcal{S}_{i} + \left( 1-\frac{n_{s}}{N} \mathcal{B}_{i}\right) \right)\, ,
\end{align}
with $\mathcal{S}_{i}$ and $\mathcal{B}_{i}$ being the probability density functions for signal and background neutrino events, respectively. 
Further, $n_{s}$ denotes the number of signal events and $N$ is the total number of recorded events.
In Ref.~\cite{IceCube:2022der} a dataset of track-like events created by muon (anti-)neutrinos with an exposure of 3186 days has been investigated.
The analysis routines provided by the IceCube collaboration~\cite{IceCubeDataset} have been modified to account for the physics scenario of this work by adjustments of the likelihood function and the inclusion of a profile likelihood ratio test. In particular, the oscillation probability of pseudo-Dirac neutrinos, cf.~Eq.\,\ref{eq:Prob}, has been incorporated in the calculation of neutrino signal events $n_{s}$ for the likelihood function in Eq.\,\ref{eq:LH}.
In addition, the likelihood function itself is added with a set of Gaussian pull terms in order to account for parameter correlations of the original analysis as well as experimental uncertainties of neutrino mixing parameters.
For the latter, we took the latest global fit results~\cite{Esteban:2020cvm,Esteban:NuFit2022} with their corresponding uncertainties. We assume normal neutrino mass ordering and for simplicity set $\delta_{\rm CP}\simeq 0$ and $\sigma_{x}\sim 10^{-10}\,\m$ - the latter is just representative of the width of the neutrino wave packet. From Eq.\,\ref{eq:Lcoh}, it is clear that this choice does not lead to additional decoherence due to wave-package separation on the Mpc-scale.
Smaller values of $\sigma_x$ would imply a smaller coherence length, and hence averaging of the spectra. This would result in a loss of sensitivity. We chose a typical value of $\sigma_x$, characteristic of charged-current production of neutrinos in AGNs, core-collapse supernovae (CCSNe) and other astrophysical objects~\cite{Kersten:2015kio}. Marginalizing over $\sigma_x$ would reduce the sensitivity slightly since smaller values of $\sigma$ would lead to decoherence. However, it is reasonable to expect that $\sigma_x$ cannot be smaller than $1$\,fm -  typical scales associated with the size of a proton.

The neutrino flux of NGC 1068 in muon and the anti-muon flavor is assumed to follow a power-law of the form 
\begin{align}
        \Phi_{\nu_{\mu} + \bar{\nu}_{\mu}} = \Phi_{0} \left( \frac{E_\nu}{\mathrm{1\,TeV}}\right)^{-\gamma}\, ,
\end{align}
with $\Phi_{0}$ being the overall flux normalization associated with an energy of $1\,\TeV$ and $\gamma$ is the spectral index.
At this point, the authors would like to emphasise that for the case of NGC 1068, the neutrino emission spectrum might be more complex~\cite{Kheirandish:2021wkm}.\footnote{For TXS 0506+056, the neutrinos were likely made in proton-photon interactions and, thus, feature a bump-like spectrum~\cite{Rodrigues:2018tku}.} 
However, in order to follow the original analysis we apply the same unbroken power-law shape as the IceCube collaboration. A different spectrum, with bump/dip-like features might reduce the sensitivity of our results. Furthermore, a contribution from muons created by tau neutrino interactions is automatically taken into account by the provided IceCube routines.
For reasons of simplicity, we leave details of the neutrino production mechanisms aside and assume a generic neutrino flavor composition of ($\nu_{e}:\nu_{\mu}:\nu_{\tau}$) = ($1:1:1$) arriving at Earth, which corresponds to a neutrino admixture of ($\nu_{e}:\nu_{\mu}:\nu_{\tau}$) = ($1:2:0$) at the source stemming from pion decay for standard neutrino oscillations. 
Deviating from these assumptions can lead to different compositions of neutrino mass states arriving at the Earth, and hence change the contribution to the survival probability in Eq.\,\ref{eq:Prob}. This will have an impact on our results and require a dedicated discussion which is beyond the scope of this work.

The corresponding number of signal events is then given by~\cite{IceCube:2021xar}
\begin{equation}\label{Eq:numev}
    \begin{aligned}
    n_{s} = t \int d\Omega \int_{0}^{\infty} &dE_{\nu}\ A_{\mathrm{eff}}(E_{\nu},\Omega)\ \Phi_{\nu}(E_{\nu}; \Phi_{0}, \gamma) \\ \times \ &P(E_{\nu}; \dm, s_{12}, s_{23}, c_{13})\, ,
\end{aligned}
\end{equation}
with the detector's lifetime $t$ and its effective area $A_{\mathrm{eff}}$ being a function of the neutrino energy $E_{\nu}$ and the solid angle $\Omega$.
The overall neutrino oscillation probability $P$ depends on the neutrino energy $E_{\nu}$, the active-sterile mass splitting $\dm$ and the standard neutrino mixing angles $s_{i}\equiv \sin \theta_{i}$ for $i=\{12,23,13\}$.
Hence, in our analysis the number of signal events depends on the six parameters: $n_{s}\equiv n_{s}(\delta m^{2}, \Phi_{0}, \gamma, s_{12}, s_{23}, c_{13})$.
Existing correlations between $\Phi_{0}$ and $\gamma$ are accounted for by reproducing the covariance matrix of the original IceCube analysis and incorporating it with a two-dimensional Gaussian pull term to the likelihood function in Eq.\,\ref{eq:LH}. 
In doing so, we also ensure that the fit preserves the original best-fit values of $\Phi_{0}$ and $\gamma$, taking into account a normalization correction due to the introduced oscillation probability in Eq.\,\ref{eq:Prob}.
Further, current knowledge of standard neutrino mixing parameters is included with individual Gaussian pull terms with uncertainties taken from Refs.~\cite{Esteban:2020cvm,Esteban:NuFit2022}.
Note that for tiny mass-squared differences $\dm \lesssim 10^{-22}$\eV$^{2}$, the effects of active-sterile mixing become negligible for the energy region of interest, such that we coincide with the usual no additional oscillation case.
In order to determine a limit on the mass-squared difference of pseudo-Dirac neutrinos, we determine a profile likelihood ratio as~\cite{Cowan:2010js}
\begin{align}\label{eq:LHratio}
    q_{\dm} = -2\log\left(\frac{\mathcal{L}(\delta m^{2}, \widehat{\widehat{\Phi}}_{0}, \widehat{\widehat{\gamma}}, \widehat{\widehat{s}}_{12}, \widehat{\widehat{s}}_{23}, \widehat{\widehat{c}}_{13}) }{\mathcal{L}(\widehat{\delta m^{2}}, \widehat{\Phi}_{0}, \widehat{\gamma}, \widehat{s}_{12}, \widehat{s}_{23}, \widehat{c}_{13}})\right)\, ,
\end{align}
where the numerator represents the likelihood function with a fixed $\dm$, while in the denominator all six quantities are tuned in the minimization procedure.
In particular, double-hatted parameters like $\widehat{\widehat{\Phi}}_{0}$ and $\widehat{\widehat{\gamma}}$ are the conditional (maximum likelihood) estimators that depend on the chosen value of $\dm$, whereas the parameters occurring in the denominator are the estimators that maximize the unconstrained likelihood function.

For each $\dm$ value under study, we determine the likelihood ratio in Eq.\,\ref{eq:LHratio} and with asymptotic sampling distributions given in Ref.~\cite{Cowan:2010js} we can assign each so-determined ratio a corresponding p-value, i.e.\ $p_{\dm} = \int_{q_{\dm, \mathrm{obs}}}^{\infty}dq_{\dm} f(q_{\dm}|\dm)$,  which we use as an exclusion criterion for the tested $\dm$ values.
Since the introduced oscillation probability in Eq.\,\ref{eq:Prob} affects the expected signal only in one direction, i.e.\ a decrease of the expected neutrino flux, we have to perform a one-sided statistical test.
Following the IceCube collaboration, we assume that the test statistic $f(q_{\dm}|\dm)$ used has approximately $\chi^{2}$-like behavior~\cite{IceCubeDataset}, except for modifications due to the one-sided test case. 
For details about the modified profile likelihood ratio and the corresponding sampling distribution, we refer to Ref.~\cite{Cowan:2010js}.
In this work, we are interested in a limit on $\dm$ at 90\% confidence level (C.L.) ($1-p_{\dm}$).
Thus, our results are determined by finding the value of $\dm$, which yields $p_{\dm}\leq 0.1$.
The whole analysis has been performed within the Anaconda/SciPy framework~\cite{AnacondaPython2022,SciPy2020}, while the iminuit package~\cite{iminuit2022} was used to minimize the likelihood function in Eq.\,\ref{eq:LH}.

\noindent\textbf{\emph{Results -- }}
\begin{figure}
\centering
    \includegraphics[width=0.5\textwidth]{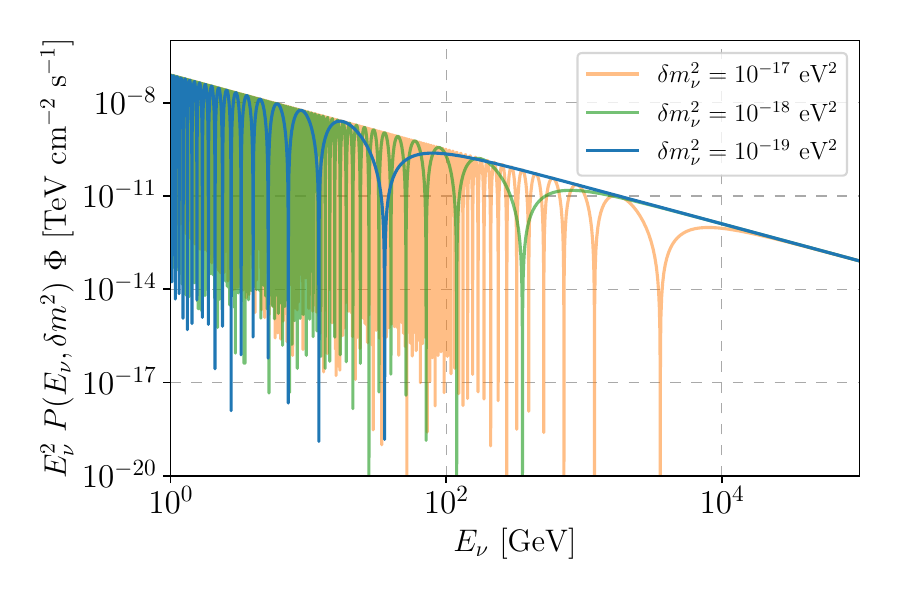}
    \caption{Neutrino flux variations due to oscillations driven by $\dm$ as a function of the neutrino true energy $E_\nu$. We rescaled the obtained neutrino flux to match the flux measured by the IceCube Collaboration~\cite{IceCube:2022der}, i.e.\ $\widehat{\Phi}_{0} \sim 5.0\times10^{-11}\,  \mathrm{TeV}^{-1} \mathrm{cm}^{-2} \mathrm{s}^{-1}$ and $\widehat{\gamma}=3.2$, and varied the pseudo-Dirac mass splitting $\dm$ in a region that is relevant for this analysis.
    }
    \label{fig:spectra}
\end{figure}
The fact that IceCube observed these neutrinos from NGC1068 implies that these did not oscillate into their sterile counterparts.
In the presence of active-sterile oscillations, the expected power-law flux will undergo spectral distortions, as shown in Fig.\,\ref{fig:spectra} for different values of $\dm=\{10^{-17}, 10^{-18}, 10^{-19}\}\,\eV^{2}$. One can see that for $\dm\gtrsim 10^{-19}\,\eV^{2}$, spectral dips are observed due to oscillation minima in the active-sterile probability. This is the crucial part in the analysis. As the value of $\dm$ changes, minima are reached accordingly when $L_{\rm osc}$ takes values close to the distance between the Earth and NGC 1068. This is the region where we expect maximum sensitivity in our results.

A profile likelihood analysis can be used to set constraints on possible values of $\dm$. In our analysis, we assume that only muon (anti-)neutrinos produced at the source are detected as muon (anti-)neutrinos, i.e.\ we set $\alpha=\beta=\mu$ in Eq.\,\ref{eq:Prob}. This implies that we only detect $\sim$ 41\% of all neutrinos. This leads to a slightly different flux normalization from the original best-fit values observed by IceCube: $\widehat{\Phi}_{0} \sim 1.2\times 10^{-10}\,  \mathrm{TeV}^{-1} \mathrm{cm}^{-2} \mathrm{s}^{-1}$, while $\gamma$ basically remains the same. In principle, the contribution of a $\nu_e$ produced at source and detected as $\nu_\mu$ is expected to be few tens of percentage of that of the original $\nu_\mu$ contribution. However, our results do not change much if we include this contribution.
\begin{figure}[!t]
    \centering
    \includegraphics[width=0.5\textwidth]{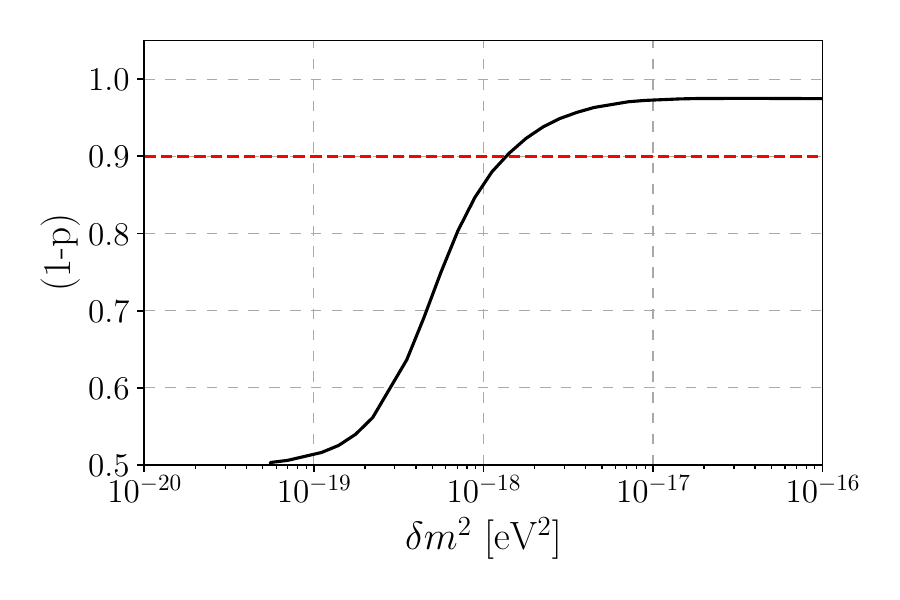}
    \caption{Exclusion for the mass-squared difference $\dm$ of pseudo-Dirac neutrinos. Confidence levels are shown for different $\delta m^{2}$ values. With the observation of neutrinos from NGC 1068, we can exclude $\dm \geq 1.4\times 10^{-18}\eV^2$ (at 90\% C.L.).}
    \label{fig:exclusion_limit}
\end{figure}
Our main result, depicted in Fig.\,\ref{fig:exclusion_limit}, shows that the data from IceCube can be used to rule out $\dm = 1.4\times 10^{-18}\eV^2$ at more than $90\%$ confidence level ($p_{\dm{}}\leq0.1$). There is some sensitivity to even lower values of $\dm$, albeit at lower confidence. This agrees with the analytical estimate in Eq.\,\ref{eq:Losc}, which shows the mass-squared difference sensitive to this oscillation length for TeV energy neutrinos. To the best of our knowledge, this is the \emph{strongest} constraint to date on the smallness of $\dm$ with such a high significance. For smaller values of $\dm$, the oscillation length becomes larger than the distance to the source, hence sensitivity is reduced.

However, note that the situation is actually more complicated. Fig.\,\ref{fig:exclusion_limit} seems to indicate that data from NGC 1068 can be used to set bounds on arbitrarily large values of $\dm$. This is certainly not feasible since, for large $\dm$, rapid variations of the oscillation probability, and hence the flux, would be averaged out by the finite energy resolution of the detector. As a result, any reduction in the number of events would be compensated by a corresponding increase in the normalisation $\Phi_0$. This is the major drawback of a counting analysis with the number of events as in Eq.\,\ref{Eq:numev}. A spectral analysis, on the other hand, would have been more sensitive to the upper limit on $\dm$.

\begin{figure}[!t]
    \centering
    \includegraphics[width=0.5\textwidth]{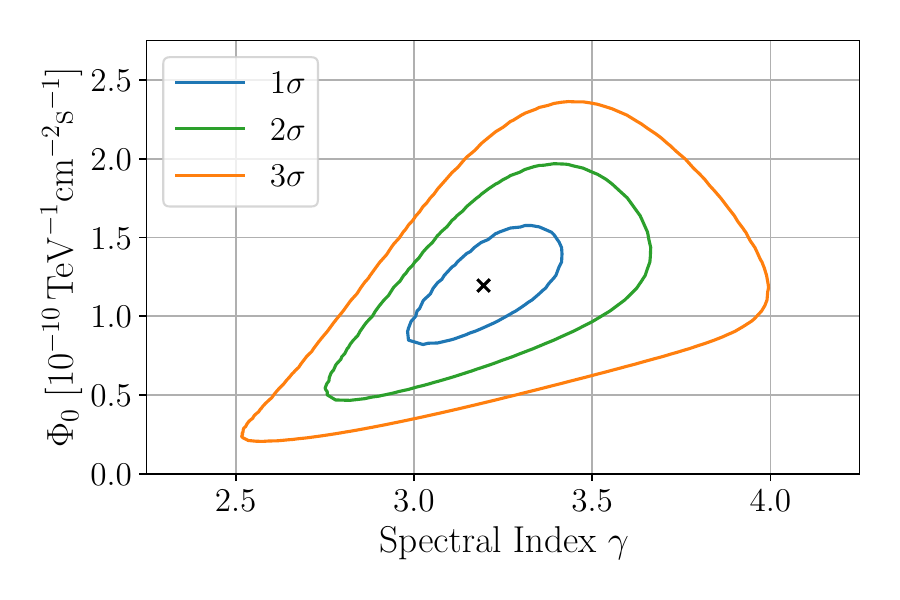}\\ ~\includegraphics[width=0.5\textwidth]{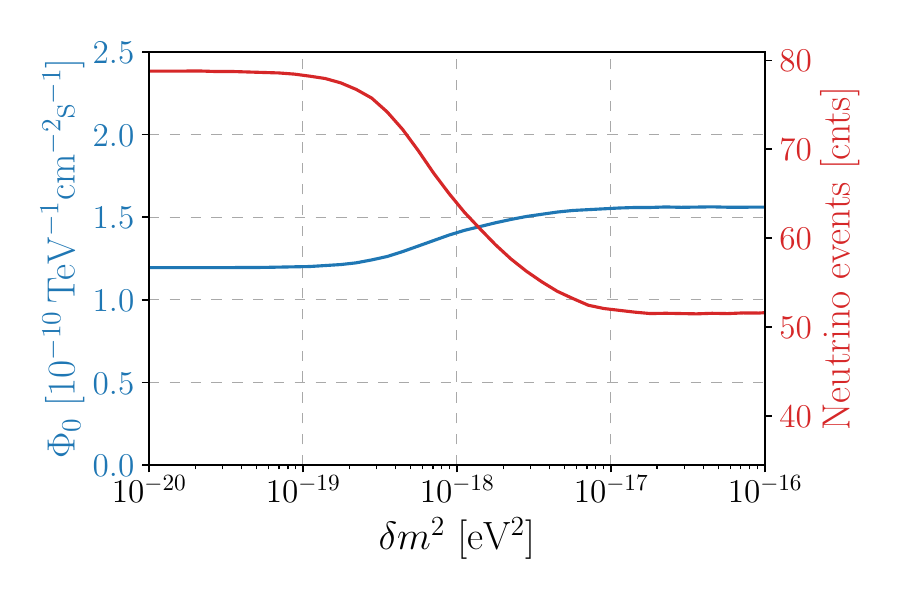}
    \caption{Top: the covariance contour to pull $\Phi_0$ and $\gamma$ to the best-fit values equivalent to the original IceCube analysis. Bottom: the $\Phi_0$ variations during the fit procedure to compensate for the reduced number of events due to oscillations.}
    \label{fig1}
\end{figure}

In this analysis, $\Phi_0$ and $\gamma$ are allowed to vary within their correlation given by the covariance matrix corresponding to the original fit. 
Hence, although a reduction in the number of events by averaged oscillations can be compensated by an increase in $\Phi_0$, the correlation between $\Phi_0$ and $\gamma$ does not allow the $\Phi_0$ to obtain values needed for a complete compensation. This is clear from the plot in Fig.\,\ref{fig1}, where we show the covariance matrix to pull $\Phi_0$ and $\gamma$ to the best-fit values corresponding to the original IceCube analysis and the $\Phi_0$ variations during the fit procedure to compensate the reduced the number of events due to oscillations. The change in $\Phi_0$ is not enough to compensate for the averaging effect for high $\delta m^2$. 

\begin{figure*}[!t]
    \centering
    \includegraphics[width=0.8\textwidth]{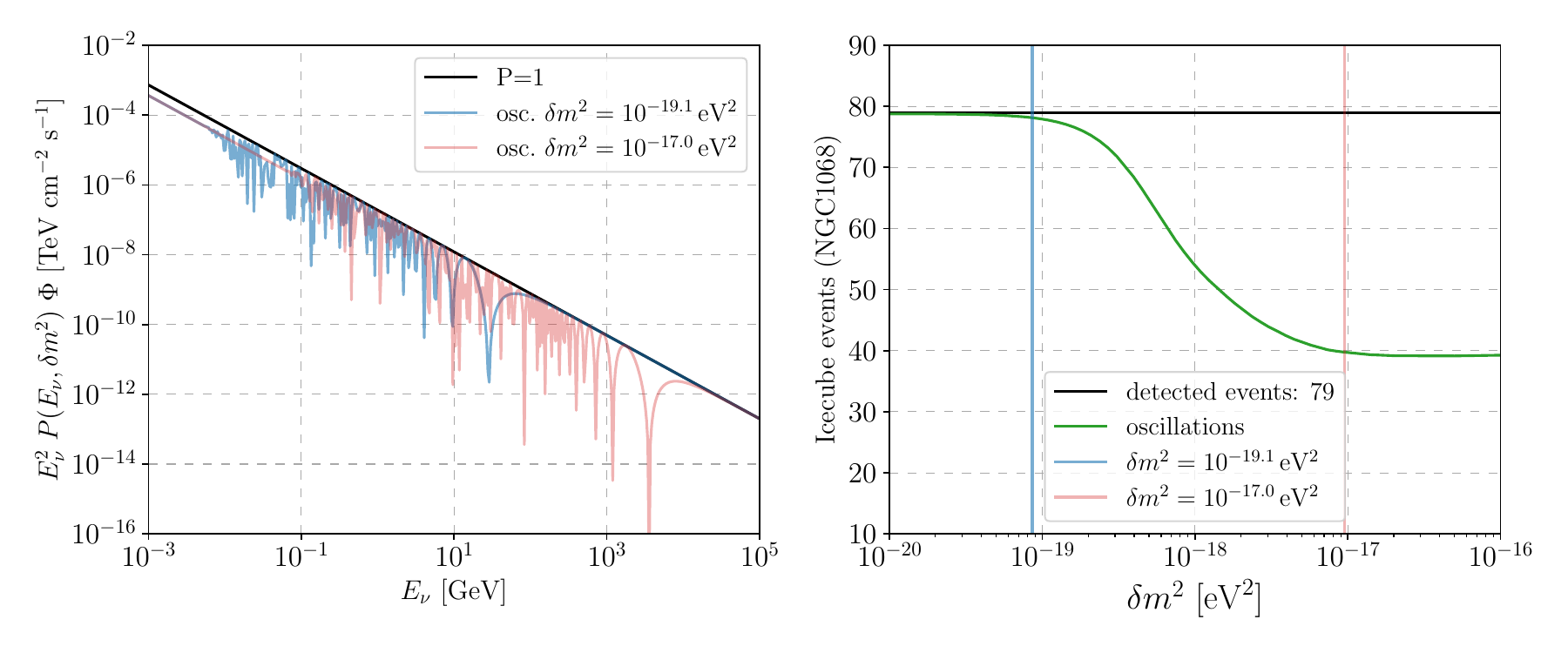}\\ 
    \caption{Left: Neutrino flux variations at the Earth due to active-sterile oscillations as a function of the neutrino true energy $E_\nu$, ( rescaled to match the flux measured by IceCuce). The standard case $(P=1)$ is compared with two different cases $\dm=10^{-19.1}\,{\rm eV}^2$ and $\dm=10^{-17}\,{\rm eV}^2$. Right: The corresponding events observed in IceCube. The vertical lines show the benchmark values of $\dm$ from where the event counts do not change by more than 1\%.}
    \label{figcomp}
\end{figure*}

To simulate the loss of sensitivity at large mass-splittings, we have performed the following test. We check for values of the mass-splitting where the survival probability, and hence the total number of events, flattens out, i.e.\ does not change the number of events below 1\%. In this regime, the effect can be mimicked by a change in the normalization of the spectra only. As a result, sensitivity would be lost in this regime. This yields the sensitivity region to be $10^{-19.1}\lesssim \delta m^2 \lesssim 10^{-17.0}$. This is explained through Fig.\,\ref{figcomp}, where the left panel shows the neutrino flux variations at the Earth for two different choices of mass-splittings within which we expect the flux variations to be relevant. The right-hand panel shows the variations of the total events, indicating the values of $\dm$ for which the event counts do not change by more than 1\%.

\begin{figure}[!h]
    \centering
\includegraphics[width=0.5\textwidth]{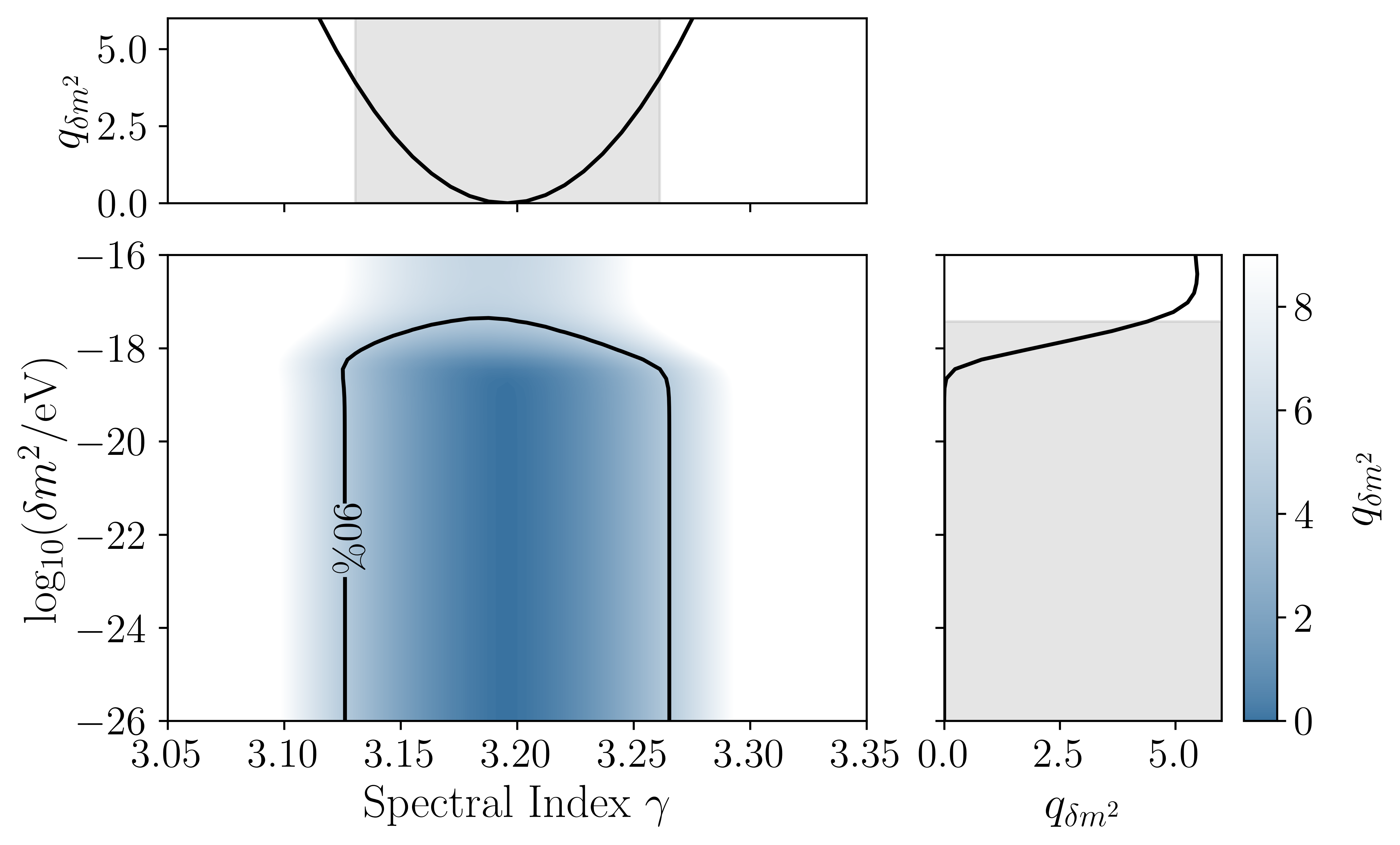}
    \caption{Parameter correlation between $\delta m^{2}$ and the spectral index for a fixed flux normalization $\Phi_{0}=1.2\times 10^{-10}\,  \mathrm{TeV}^{-1} \mathrm{cm}^{-2} \mathrm{s}^{-1}$. The black lines denote the 90\% CL contours for these parameters.}
    \label{fig:mass_vs_gamma}
\end{figure}

Fig.\,\ref{fig:mass_vs_gamma} shows the correlation between $\dm$ and $\gamma$ for a fixed flux normalization $\Phi_{0}=1.2\times 10^{-10}\,  \mathrm{TeV}^{-1} \mathrm{cm}^{-2} \mathrm{s}^{-1}$. Analogous to our determined limits, we indicate the 90\% C.L. contour for both parameters. Correlations between $\Phi_{0}$ and $\gamma$ are considered via an appropriate pull term in the applied likelihood function. The figure indicates that the data is clearly consistent with the ``no-oscillation" hypothesis, which corresponds physically to oscillations with longer baselines. This shows that the addition of these oscillations does not help improve the fit, rather the data can be used to put constraints on these oscillations.

Finally, note that uncertainties of the source distance have not been accounted for in this analysis. Since we are varying $\delta m^2$, there will always be a certain range of $\delta m^2$ for which the neutrino oscillation probability will undergo a few oscillation cycles. This will not get averaged out due to the uncertainty in the distance.
Of course, this might lead to a slight reduction in sensitivity, but that is subdominant with respect to the effect of the other uncertainties like the determination of the neutrino energy.

A similar argument can be used to put bounds on $\dm$ from observations of a neutrino event of energy $E_\nu\sim 290\, \TeV$ from the blazar TXS 0506 + 056 at a redshift of $z=0.33\pm 0.0010$, corresponding to a distance $d=1.3\,\Gpc$~\cite{IceCube:2018cha,IceCube:2018dnn}. A naive estimate yields
\begin{equation} \label{eq:Losc_blazar}
    L_{\rm osc} 
    \approx 1.3\,\Gpc\left(\frac{E_\nu}{290\,\TeV}\right)\left(\frac{10^{-17}\,\eV^2}{\delta m^2}\right)\,,
\end{equation}
which is stronger due to the larger energy of the neutrino. Additionally, IceCube later reported a series of $13\pm5$ events from the same direction. These observations can be used, in principle, for a similar study to constrain values of $\dm$.

\noindent\textbf{\emph{Conclusion -- }} 
Detection of high-energy neutrino events by IceCube has opened up new frontiers in multi-messenger astronomy. In particular, the naturally long baseline available to the neutrinos, coupled with their high energies, allows them to be the harbingers of new exotic physics, otherwise inaccessible to mankind. In this work, we used the IceCube observation of $\sim 79$ TeV-sh neutrino events coming from the direction of the active galactic nuclei NGC 1068 to probe the extent of lepton number violation in the Standard Model.

A soft lepton number violation can be manifested through active-sterile neutrino oscillations over astronomical baselines inversely proportional to the mass-squared difference $(\dm)$ between active and sterile neutrinos. These oscillations, if present, will lead to a distortion of the event spectra, and a reduction in the number of neutrinos observed. 

This simple yet powerful idea can be utilized to probe the strength of lepton number violation, characterised by the magnitude of $\dm$. The identification of the source of these neutrinos, and precise measurement of their energies, allowed us to precisely calculate the oscillation lengths associated with $\dm$, and the corresponding spectral modifications. We used the analysis of the IceCube collaboration, outlined in~Ref.~\cite{IceCube:2022der}, to rule out $\dm = 1.4\times 10^{-18}\eV^2$ at more than $90\%$ confidence level, using the total number of events observed. We simulated the loss of sensitivity at large mass-squared splitting by checking where the  
number of events does not change by more than 1\% when the mass-splitting is changed. This yields the sensitivity region as 
$10^{-19.1}\lesssim \delta m^2 \lesssim 10^{-17.0}\,{\rm eV}^2$.

Parameter correlations of the original IceCube analysis as well as uncertainties of the neutrino mixing angle have been properly treated via Gaussian pull terms. The result obtained is one of the strongest bounds on the extent of lepton number violation from experiments.
However, we would like to point out that the performed investigation depends crucially on the assumptions of the neutrino energy spectrum as well as their flavor composition.
We leave a detailed study on the impact of various flux assumptions as well as a more involved sterile neutrino sector, i.e. different $\dm_{i}$, for future analyses.

These extreme new physics scenarios are clearly inaccessible to any terrestrial experiments. The observations of these point-sources of astronomical neutrinos like NGC 1068 and TXS 0506+056 clearly pave the way for new laboratories for neutrino physics.

\noindent\textbf{\emph{Acknowledgements --}} We thank Yuber Perez-Gonzalez for discussions and collaborations in the early phase of the project, and Sergio Palomares Ruiz for useful comments on the draft. We would also like to thank the IceCube Collaboration for providing their datasets and analysis routines to the public, which set the foundation for the analysis performed in this work.

\bibliographystyle{apsrev4-1}
\bibliography{references.bib}
\end{document}